# Back with Weight: Revisiting Very Heavy Ions for Precision Radiotherapy

Jeannette Jansen[1], Olga Sokol[1], Yolanda Prezado[2,3,4,5] and Marco Durante[1,6,7*]

[1] GSI Helmholtzzentrum für Schwerionenforschung, Biophysics Department, Planckstraße 1, 64291 Darmstadt, Germany

[2] New Approaches in Radiotherapy Lab, Center for Research in Molecular Medicine and Chronic Diseases (CIMUS), Instituto de investigación Sanitaria de Santiago de Compostela (IDIS), University of Santiago de Compostela, Santiago de Compostela, A Coruña, Spain

[3] Oportunius Program, Galician Agency of Innovation (GAIN), Xunta de Galicia, Santiago de Compostela, A Coruña, Spain

[4] Institut Curie, Université PSL, CNRS UMR3347, Inserm U1021, Signalisation Radiobiologie et Cancer, Orsay, France

[5] Université Paris-Saclay, CNRS UMR3347, Inserm U1021, Signalisation Radiobiologie et Cancer, Orsay, France

[6] Institute of Condensed Matter Physics, Technische Universität Darmstadt, Hochschulstraße 6-8, 64289 Darmstadt, Germany

[7] Department of Physics „Ettore Pancini", University Federico II, Monte S. Angelo, Naples, Italy

*Corresponding author: M.Durante@gsi.de

# Abstract

Accelerated charged particles offer significant physical advantages over X-rays in radiotherapy. In addition to their superior depth-dose distribution, heavy ions provide notable biological benefits compared to protons. Specifically, at therapeutic energies, very heavy ions are expected to exhibit high relative biological effectiveness and a low oxygen enhancement ratio, making them potentially ideal for treating radioresistant tumors. Over fifty years ago, the Lawrence Berkeley Laboratory started a clinical trial to treat cancer patients with particles ranging from helium to argon. However, treatments with ions of atomic number greater than 10 proved highly toxic to normal tissue. Most patients were ultimately treated with helium ions, whose biological effects closely resemble those of protons. In recent years, novel strategies have emerged that can reduce normal tissue toxicity in radiotherapy while preserving tumor control. These advancements open the possibility of revisiting the clinical use of heavier ions. In this paper, we propose that neon ions may serve as an effective modality for treating resistant, hypoxic tumors. Their associated toxicity may be mitigated by employing approaches such as ultra-high dose rate (FLASH) or spatially fractionated radiotherapy. Current plans and prospects for the clinical application of neon ions will be discussed.

**INTRODUCTION**

Charged particle therapy is widely regarded as the most advanced form of radiotherapy currently available *(1)*. The concept was first proposed by Robert R. Wilson, a student of Ernest Orlando Lawrence at the University of California, Berkeley. Wilson recognized that protons, due to their favorable depth-dose distribution (Bragg peak) could offer significant dosimetric advantages over conventional X-rays *(2)*. By the end of 2024, more than 400,000 patients had been treated with particle therapy for various types of cancer *(3)*. Approximately 85% of these patients received proton therapy, which remains the only modality available at nearly 90% of the particle therapy facilities in operation as of 2025 *(4)*.

Nevertheless, in his seminal paper, Wilson also suggested the potential therapeutic benefits of using ions heavier than protons. This idea was further developed at the Lawrence Berkeley National Laboratory (LBNL) in California by Cornelius A. Tobias *(5)*. During the 1970s, LBNL launched an extensive series of preclinical studies on the radiobiology of charged particles, revealing that heavy ions offered several biological advantages in addition to the physical benefits already demonstrated by protons *(6)*. These advantages stem from the ability of heavy ions to achieve linear energy transfer (LET) values comparable to those of $\alpha$-particles or neutrons within the therapeutic energy range *(7)*. At such high-LET values, direct DNA damage dominates *(8)*, leading to complex lesions that are difficult for cells to repair.

Pioneering studies by Eleanor A. Blakely and colleagues at LBNL *(9)* showed that heavy ions significantly increased the relative biological effectiveness (RBE) and reduced the oxygen enhancement ratio (OER), particularly at LET values around 100–200 keV/µm (Fig. 1). The reduction in OER was especially important, as tumor hypoxia was already recognized as a negative prognostic factor *(10)* that renders cancers resistant to radiotherapy and chemotherapy *(11)*. Additional biological advantages of high-LET particles include reduced sensitivity to dose fractionation and to the cell-cycle phase at the time of irradiation *(12)*. More recently, further benefits of densely ionizing radiation have been identified, such as decreased tumour angiogenesis *(13)*, suppression of tumour cell migration and invasiveness *(14)*, and enhanced anti-tumor immune response *(15–18)*.

**THE BERKELY CLINICAL TRIAL**

Meanwhile, Ernest Lawrence's brother, the physician John, started to treat cancer patients at LBNL with light ions with the support of Tobias already in the 60's (19). Tobias then expanded this program, launching the first pilot trial of heavy ion therapy that treated over 1,300 cancer patients with different charged particles (20). The original idea was to use ions from light to heavy: $^4$He, $^{12}$C, $^{20}$Ne, $^{28}$Si, and $^{40}$Ar (21). The use of very heavy ions was justified by the fact that the OER drops to ~1 only at very high-LET values (Fig. 1) (9,22,23). To treat a solid tumor, whose diameter is in the order of several cm, it is necessary to overlap ion beams of different energies to create a spread-out-Bragg-peak (SOBP). The dose-averaged LET (LETd) in the SOBP is much lower than the LET of a monoenergetic beam, particularly for large tumors (Fig. 2). For this very reason, Tobias hoped to use very heavy ions against very hypoxic and radioresistant tumors, such as glioblastoma multiforme (GBM).

Unfortunately, for very heavy ions the LET is high already in the entrance channel of the Bragg curve, making them very toxic. For instance, argon ions have an LET around 100 keV/$\mu$m, around the peak of the RBE vs. LET curve (Fig. 1), already at 200 MeV/n i.e. when traversing normal tissue. Eventually, in the LBNL trials (1975-1992) the majority of patients were treated with He-ions and only 433 with heavier ions (24).

When the LBNL Bevalac accelerator was shut down in 1992, Hirohiko Tsujii decided to run a heavy ion program at the National Institute of Radiological Science (now National Institute for Quantum Science Technology; QST) in Chiba (Japan) focusing on carbon ions only, a particle that presented a low-LET in the entrance channel and a relatively high-LET in the SOBP (Fig. 2). Over 16,000 patients have been treated in Chiba since 1994, with excellent results (25). Carbon ions were also selected by Gerhard Kraft at the GSI Helmholtz Center for Heavy Ion Research in Darmstadt (Germany) for the first heavy ion trial in Europe (1997-2008) (26). Carbon is currently used in 17 clinical centers in Asia and Europe for treatment of different types of solid malignancies (4). A carbon ion therapy center is currently under construction at the Mayo Clinic in Jacksonville, Florida (27). This facility will mark the return of heavy ion therapy to the United States after more than three decades.

**HEAVIER THAN CARBON?**

While the results of the C-ion trials are generally considered good, they do not seem to have led to the breakthrough that was originally expected by Tobias and his co-workers at LBNL. The reason is probably clear if we compare Fig. 1 and Fig. 2: the LET of the C-ions in the SOBP is simply not high enough to reach the peak of the RBE, and certainly not to overcome hypoxia. There is a general feeling in the particle therapy community that C-ion therapy cannot be really defined "high-LET therapy" (28).

The radiobiology data are now supported by clinical evidence. Local recurrence has been correlated with low LETd values in patients treated with carbon ions for pancreatic cancer (29,30), sacral chordoma (31) and chondrosarcoma (32). Several strategies have been proposed to increase the LETd in C-ion therapy, including simultaneous integrated boost (33), arc therapy (34) and LET painting (35,36). Nevertheless, the basic physics suggest that the simplest solution would be to use heavier ions, either alone or in combination with lighter particles (multi-ion therapy) (37,38). At the moment, the Heidelberg Ion Therapy (HIT) center in Germany is planning to use $^{16}$O-ions for radioresistant tumors (39), whereas QST in Japan is considering the use of both oxygen and neon ions in combination with lighter ions (protons, helium or carbon) in multi-ion therapy (40,41). As yet, only one patient has been treat at QST for a sarcoma using carbon and oxygen ions. This single patient is the first to be treated with a ion heavier than carbon by the end of the LBNL trial in the USA.

What could be the ideal "heavier ions" for treatment of very radioresistant tumors? We propose that neon may be the right candidate. As shown in Fig. 2, neon will still maintain an acceptable low-LET in the entrance channel, but is going to have LETd>100 keV/μm all over the SOBP. This point is particularly relevant because hypoxia exhibits significant spatial heterogeneity (42,43), and it cannot be assumed that the tumor's center corresponds to the hypoxic core targeted for boosting (44,45). In GBM, it has been recently shown that spatially distinct hypoxic niches are associated to tumour-myeloid cells immunosuppressive interactions (46–48) and are therefore the most critical structures for targeting therapy. Only a uniform high-LET in the target region can guarantee that these niches can be effectively destroyed, thus activating immune response against the tumour. A comparison of C- and Ne-ions for the treatment of a human GBM is shown in Fig. 3. The treatment plan, calculated with TRiP98 (49), shows that at the same target coverage in RBE-weighted dose, the $LET_d > 100$ keV/$\mu$m for Ne-ions in almost the whole target volume, while for carbon it remains consistently below.

In addition, neon has good physical characteristics compared to heavier ions, where the Bragg curve is bended by nuclear fragmentation. In Fig. 4 we show the Bragg curves for H-, C-, Ne- and Ar-ions at the same range (18 cm in water) normalized either at the entrance dose or at the Bragg peak dose. Argon is clearly the worst ion in terms of depth dose distribution. Moreover, this range requires an energy of 520 MeV/n for $^{40}$Ar-ions, which cannot be reached at medical synchrotrons. In fact, synchrotrons in operation for cancer therapy accelerate carbon ions up to 430 MeV/n, corresponding to a rigidity of 6.6 Tm. At this rigidity, Ne-ions can be accelerated up to a range in water around 18 cm, sufficient to treat brain and head-and-neck tumors, while for argon ions the maximum achievable range would be 8.3 cm only, making it suitable for shallow tumors only.

Finally, neon has been used in the LBNL clinical trial with promising results. A total of 239 patients were treated with Ne-ions from 1979 to 1988 at LBNL with a minimum dose of 1 Gy (50). In patients treated for salivary gland carcinoma, paranasal sinus tumour, soft tissue or bone sarcomas, locally advanced prostate cancer and biliary tract carcinoma, local control appeared better than the standard of care with X-rays. For tumors characterized by dismal prognosis, such as pancreatic cancer or GBM, outcome was, however, similar to conventional treatments. In the initial GBM trial, 8 patients were treated with Ne-ion doses ranging 17-22 Gy and median survival was only 6 months. In a subsequent randomized protocol, 15 GBM patients were treated with 20 or 25 Gy Ne-ions in 4 weeks (51). Even if the median survival remained disappointing (13 and 14 months in the low- and high-groups, respectively), there was evidence of improvement with dose and 3 patients died after nearly 2 years for other causes and no evidence of residual tumour. The doses used are still relatively low compared to the current standard of care for X-rays is 60 Gy in 30 fractions, but RBE is uncertain and a further dose escalation would be problematic because of the risk of toxicity. It is noticeable that whilst, as discussed above, low-LETd volumes are associated to local recurrence, high-LETd volumes in the normal tissue can be associated to toxicity, such as sacral insufficiency fractures after C-ion radiotherapy for sacral chordoma patients (52). A mitigation strategy for heavy ion radiotherapy is certainly necessary.

Nevertheless, the LBNL experience shows that Ne-ion therapy is feasible, safe and can be effective is dose escalation if possible.

**REDUCING TOXICTY**

Since toxicity was the main reason to discontinue the use of very heavy ions in therapy, we have to ask whether modern approaches can spare better normal tissue to the point that heavy ions such as neon can be reintroduced. In fact, in recent years a number of physical and biological developments in the field of radiation oncology led to new techniques potentially able to spare normal tissue in a much more effective way that so far possible (53).

For instance, in the PULSAR approach (54) a high dose delivered by stereotactic ablative radiotherapy is spaced by long intervals (weeks or even months) when the patient is kept under strict surveillance and eventually treated with immunotherapy before a second fraction ("pulse"). PULSAR needs accurate predictive mathematical or AI models for optimizing the treatment in a personalized way (55,56), but is meanwhile entering already in clinical trials (57,58). Heavy ions can be excellent for PULSAR because they are potentially more immunogenic than X-rays (59) and the long intervals between the fractions will reduce toxicity.

A second interesting example for heavy ion therapy is the partial tumour irradiation targeting hypoxic segment (PATHY) method, where the idea is to target the hypoxic part of the tumour with a high dose while sparing the immune islands in the tumor microenvironment to allow a strong anti-tumor immune response (60). The idea has been tested in the clinics with both X-rays (61) and carbon ions (62) and gave amazing results especially for bulky, unresectable large tumors (63). Again, since only a fraction of the tumour is irradiated and PATHY relies on a strong immune response, heavy ions could be an ideal tool for this approach. The multi-ion therapy proposed at QST (40) is actually aiming at increasing the LET in the central part of the tumour using ions as heavy as oxygen (Fig. 5) or neon (41), in an approach that is similar to a heavy ion PATHY.

In this review we will now focus on two major techniques for sparing normal tissue: spatially fractionated radiotherapy (SFRT), with particular regard to minibeam radiotherapy (MBRT); and ultra-high dose rate (FLASH) radiotherapy.

**SPATIALLY FRACTIONATED RADIOTHERAPY**

The term SFRT includes a number of distinct techniques such as GRID (64), lattice (65), MBRT (66) or microbeam (67) radiotherapy. These techniques share the old observation that normal tissue has much better recovery when the beam targeting the tumour is spatially divided using, as simplest approach, a grid (68). The unirradiated parts of the normal tissue will repopulate quickly the sectors destroyed by radiation, leading to reduced toxicity (69). SFRT comes with a concern for tumour control, but mounting evidence shows that non-uniform target dose distributions do not jeopardize tumour control (70–74). While all these techniques are now slowly entering in the clinical practice with X-rays or protons, a first mouse study of MBRT using Ne-ions (75) has been successfully performed at the HIMAC accelerator at QST in Japan (Fig. 6), paving the way to future applications of MBRT with very heavy ions.

The biological foundations of SFRT extend far beyond geometric dose modulation. Although initially conceived from empirical observations of superior normal tissue tolerance to heterogeneous irradiation, growing evidence reveals that SFRT elicits complex biophysical and immunological mechanisms involving tissue regeneration (dose-volume effects), vascular modulation, intercellular signaling (bystander-like effects), and immune activation (74).

Stem cell migration and proliferation within low-dose (valley) regions are thought to be central to SFRT's ability to spare normal tissues (76). Surviving clonogenic or stem cells in these valleys may migrate toward high-dose (peak) zones, promoting tissue repair and regeneration. Evidence from animal studies supports this idea, showing rapid skin and germ cell recovery after MBRT due to surviving valley cells (77–79).

In SFRT, vascular effects could be central to the biological response to irradiation. High doses used in GRID and lattice radiotherapy (80,81) induce endothelial apoptosis via the acid sphingomyelinase (ASMase) pathway, leading to ceramide-mediated vascular collapse and tumor cell death—similar to high-dose radiosurgery, where tumor control arises from vascular destruction. In contrast, microbeam radiation therapy differentially affects tumor and normal vessels: immature tumor vasculature is preferentially damaged, whereas mature vessels in healthy tissue are largely preserved, maintaining normal perfusion (82). Comparisons between microbeam and broad beam irradiation show reduced tumor perfusion and altered vascular integrity (83), though some studies report vascular repair and reorganization after microbeam and MBRT (84). This partial vascular preservation may support immune cell infiltration, a key factor in the anti-tumor response elicited by SFRT.

The immune response has emerged as another fundamental mechanism underlying SFRT's therapeutic efficacy. Preclinical evidence shows that T cells, particularly CD8+ cytotoxic lymphocytes, are essential for tumor control after SFRT, with SFRT inducing rapid and robust immune infiltration compared to conventional radiotherapy (85,86). One potential explanation is that the high-dose peaks in SFRT promote immunogenic tumor cell death and the release of neoantigens, enhancing antigen presentation and systemic immune activation, while immune cells within the low-dose valleys are preserved and remain functional. These effects contribute to both local and distant (abscopal) responses, as demonstrated in studies where SFRT reduced growth in both irradiated and non-irradiated tumors (81). Moreover, combining SFRT with immune checkpoint inhibitors such as anti-CTLA-4 or anti-PD-1 has produced synergistic outcomes (87), indicating its potential as an immunomodulatory platform that reprograms the tumor microenvironment toward an immune-active phenotype.

The generation of reactive oxygen species (ROS) during SFRT has been hypothesized as an important biochemical factor influencing its biological effects (88). Beyond their well-known role in radiation-induced DNA damage, ROS such as hydrogen peroxide ($H_2O_2$) also participate in cell signaling and modulation of the tumor microenvironment. It has been proposed that $H_2O_2$ formed in the high-dose peak regions can diffuse into the adjacent low-dose valleys, creating a more uniform oxidative distribution that may help explain the observed tumor responses in microbeam and MBRT (88,89). Recent Monte Carlo simulations (90) have revealed that ROS generation differs between beam types and spatial dose regions. In carbon MBRT, valley zones showed higher yields of hydroxyl radicals and aqueous electrons but lower levels of $H_2O_2$ compared to peaks. Interestingly, reduced extracellular $H_2O_2$ concentrations have been associated with enhanced anti-tumor immune activity (91), suggesting that the spatial modulation of oxidative stress in SFRT may contribute not only to direct cytotoxicity but also to the stimulation of immune-mediated tumor control.

To date there are no indications that the above described mechanisms would differ depending on the dose-rate or LET, other than a potential synergistic enhancement of immune effects when SFRT is combined with heavy ions, such as Neon.

**FLASH**

Since the first demonstration that irradiation with ultra-high dose rate (>40 Gy/s) electrons was able to reduce fibrosis in mice whilst maintaining tumour control (92), studies on FLASH using different radiation qualities have rapidly increased and are entering in the clinics (93). While ultra-high dose rate is technologically easier to be achieved in high-intensity particle accelerators such as electron linacs and proton cyclotrons (94), at GSI we have shown that FLASH conditions can be reached also in synchrotrons, when all particles needed to deliver the dose are compressed in a single synchrotron spill (95). The FLASH beamline in the Cave A of the GSI synchrotron SIS18 is shown in Fig. 7.

The most important part of the beamline is the 3D range modulator (96–98) that allows ultrafast dose delivery for simple geometrical target volumes or arbitrary clinical contours and creates highly conformal 3D dose distributions starting from a monoenergetic beam scanned in 2D.This is essential for 3D conformal Bragg peak FLASH irradiation, because active energy changes are too slow to irradiate the volume in <1 s. Using this beamline, the FLASH effect has been then demonstrated in a mouse model both in the entrance channel (99) and in the SOBP (100) of a therapeutic carbon ion beam. Feasibility of FLASH irradiation with oxygen ions has recently been shown at HIT, but no toxicity endpoints were measured (101).

Will the FLASH effect be observed also at very high-LET? This is actually unclear, because a mechanistic explanation of the FLASH effect is still missing. The original hypothesis to explain the FLASH effect was the oxygen depletion (102), i.e. the generation of transient anoxia, boosting radioresistance, when irradiation time is shorter than oxygen diffusion velocity. However, chemical models show that oxygen depletion cannot explain the FLASH effect at the dose commonly used in the experiments (103). Direct measurements of radiation-induced changes of intracellular oxygen levels also show that the reached levels of oxygen depletion are too small to explain the FLASH effect based on radiation-induced hypoxia (104–106). However, a reduction, or a shift in residual oxygen-derived radicals, such as $\bullet OH$ and $H_2O_2$, has been reported, potentially leading to less damage. While oxygen depletion seems to be unable to explain the FLASH effect, oxygen still plays a key role in the FLASH-effect, as several studies have observed that the protective effect of FLASH only occurs in certain oxygen regimes, including in vitro experiments with C-ions (107). This fact makes it even more interesting to test what will happen at higher LET, where the OER is expected to be ~1, and oxygen plays potentially less of a role.

Several other models have been proposed to explain the FLASH effect but none of them is currently universally accepted in the community (108,109). A summary of the current models is provided in Table 1. Interestingly, all of them are LET-dependent (95). Models based on oxygen depletion predict that the FLASH effect should decrease with LET, and eventually disappear at LET>100 keV/$\mu$m (110–112), but other models contend that with high-LET the sparing effect will still remain unchanged (113) and the tumour control may be even enhanced at ultra-high dose rate (114,115).

*Table 1. Summary of the current mechanistic models published to explain the FLASH effect.*

| # | MODEL / HYPOTHESIS | Dose-rate dependence | LET-dependence |
|---|---|---|---|
| 1 | **Oxygen depletion** (leads to radiation-induced hypoxia) | Less depletion at higher dose rates in vitro (104,106,116,117). Similar depletion in vivo (118). | Less depletion at higher LET. Difference between FLASH and conventional dose-rate decreases at high LET (112). |
| 2 | **Radical-radical recombination** (leads to less secondary radicals contributing to DNA damage) | Less secondary radicals at FLASH due to closer temporal proximity (119–121). Notably, while models predict higher $H_2O_2$ production with FLASH, experimental data found a decrease of this molecule at ultra-high-dose rate. | Increased $H_2O_2$ at higher LET (122). $O_2$-independent. |
| 3 | **Lipid peroxidation (LPO)** | Less LPO at FLASH, tissue- and O2- dependent (123,124). | Less LPO at high LET (125). |
| 4 | **Vasculature-effects, hemolysis** | Vascular dilatation observed at conventional dose-rate but not in FLASH, microvasculature is preserved in brain (126). | Stronger vascular damage at higher LET (127). |

| 5 | **Immune system** | No dose-rate dependence observed (128). Less lymphopenia with FLASH due to shorter exposure time (129). Faster recovery of T-lymphocytes in FLASH (130). | Enhanced immune response at high-LET (131). Reduced lymphopenia due to volume effect (132). Observed reduction of metastasis with C-ion FLASH (99). |

Another promising mechanism to explain the FLASH effect could be lipid peroxidation (LPO). Cells and nuclei possess a bi-layered lipid membrane, which gets damaged during radiation. Experiments with phosphatidylcholine liposomes revealed a strong dose-rate dependence of LPO (123,124), as a key reaction within the LPO is several orders of magnitudes slower than the other radiation-mitigated chemical reactions. This leads to a "saturation" when a lot of radicals are produced at the same time, as happens in FLASH. As a result, less LPO is observed at high dose rates. In addition, lipid content varies across different tissues and tumors, and could be contributing to the fact that tumors' response to FLASH remains typically unaltered and only the normal tissue is affected. In the context of high-LET radiation, according to studies from the 90s, LPO is expected to be smaller for increased LET, already at conventional dose rates (125). Similar to the oxygen depletion mechanism (which also plays into LPO), one could expect less difference in LPO between FLASH and conventional dose rates, but this remains to be experimentally tested, especially in vivo.

In addition to the radiation chemistry-driven mechanisms, several other mechanisms were elaborated, such as a change in vascularization, or an altered immune response. From a physics perspective, it is likely that the changes in these "biological endpoints" in fact start from an altered radiation chemistry. An exception in this is the immune response, which seems to be activated in a different fashion, both via FLASH and high-LET radiation: the more clustered DNA damage, caused by high-LET, promotes typically a more specific anti-cancer immune response (133). A higher recovery of T lymphocytes was reported after total-body irradiation of mice in FLASH conditions compared to conventional dose-rate (130). Other studies, however, found no difference in immune response for different dose rates (128). Regarding metastasis, high-LET radiation has shown promise to cause less cell migration and epithelial-to mesenchymal transitions in vitro (134) corresponding to less metastasis in vivo, al already discussed previously in this review.

FLASH high-LET experiments are therefore warranted to test the existing FLASH models as well as to expand into further applications (93,135,136).

**THE HADRONMBRT PROJECT**

The combination of MBRT and very heavy ions, such as neon, was first proposed by Peucelle *et al.* (137) as a novel strategy to mitigate the toxicity historically associated with these ions, while exploiting their high efficiency in killing hypoxic tumor cells. Early Monte Carlo (MC) simulations indicated the potential of heavy-ion MBRT for normal tissue sparing and identified neon as offering the most favorable dosimetric profile among the ions studied (137,138). These studies reported extremely high peak-to-valley dose ratio values, very narrow penumbras, and low valley doses in both the proximal beam path and the fragmentation tail, thereby enhancing healthy tissue sparing near the beam entrance. Although the production of secondary nuclear fragments increases with atomic number, MC simulations indicated that their contribution to the valley dose becomes significant only at the Bragg peak (tumor) position (137,138). These promising findings strongly motivated subsequent experimental studies.

The first implementation of neon MBRT took place in 2018 at QST, in collaboration with the Department of Charged Particle Therapy Research, National Institute of Radiological Sciences (NIRS). The initial dosimetric characterization and biological evaluation of Ne minibeams were conducted by Prezado et al. (75) at the Heavy Ion Medical Accelerator (HIMAC) at QST. In vivo experiments revealed that mouse legs irradiated with 20 Gy of neon broad beams developed severe damage, including necrosis, whereas those treated with neon minibeams exhibited only mild, reversible erythema (Fig. 6). These results demonstrated a substantial improvement in normal tissue sparing with MBRT. Building on these findings, additional experimental campaigns—partially funded by the CNRS International Emerging Actions program—confirmed the healthy tissue sparing potential of Ne-ions minibeam radiotherapy (NeMBRT) and suggested promising tumor control efficacy (unpublished).

In parallel, González-Vegas et al. (139) provided mechanistic insights by combining NeMBRT with synchrotron radiation-based Fourier transform infrared microspectroscopy. Normal and tumor cell lines were irradiated at HIMAC with either neon broad- or mini-beam radiation. Principal component analysis of the infrared spectral signatures revealed distinct biochemical responses between the two modalities. Normal cells exposed to NeMBRT recovered faster and exhibited fewer persistent molecular alterations, while tumor cells maintained strong radiobiological responses. These observations further support the selective tissue-sparing properties and immunogenic potential of NeMBRT, providing molecular-level evidence for its continued preclinical development.

Collectively, these studies indicate that combining neon ions with MBRT offers a promising strategy to overcome the normal tissue toxicity that has historically limited the clinical use of very heavy ions. Beyond its tissue-sparing effects, MBRT has been shown to modulate the tumor microenvironment and induce durable antitumor immune memory—often referred to as the "vaccine effect" (86). Given that heavy ions have also been linked to stronger immune activation compared to photons and protons (15–18), NeMBRT may achieve synergistic immunomodulatory effects. Moreover, recent findings showing a reduced oxygen dependence in proton MBRT (140) for both normal tissue protection and tumor control could complement the inherently low OER of neon ions.

For these reasons, NeMBRT emerges as a highly promising approach to widen the therapeutic index for the treatment of hypoxic and radioresistant tumors, potentially restoring clinical interest in very heavy ion therapy. By combining the intrinsic advantages of neon ions— high LET, low OER, and strong hypoxic cell-killing capacity—with the spatial fractionation benefits of MBRT, this strategy addresses two of the major limitations in radiation oncology: tumor radioresistance and normal tissue toxicity. Future perspectives include a comprehensive immune profiling to elucidate the mechanisms underlying NeMBRT's immunogenic potential, the investigation of oxygen significance within both normal and tumor tissues to exploit the synergy between neon's low OER and MBRT's reduced oxygen dependence and long-term tumor control and toxicity studies in relevant preclinical models to establish optimal beam parameters and dosing strategies.

If confirmed in translational and early-phase clinical studies, NeMBRT could represent a paradigm shift in the use of very heavy ions, transforming them from a historically abandoned modality into a precision radiotherapy tool capable of safely tackling some of the most challenging tumors.

**THE HI-FLASH PROJECT**

Based on the idea of using ultra-high dose rate to reduce toxicity of heavy ions, HI-FLASH is a new project based at GSI in Darmstadt and supported by the European Research Council (ERC) whose goal is to treat an animal brain tumour with high-LET neon ions in FLASH conditions. The focus will be on GBM which is, as described before, often considered for heavy ion therapy because of its high hypoxia. There are about 250,000 new GBM cases per year in the world and incidence is increasing (141). The current standard-of-care involves surgical resection followed by RT (60 Gy in 30 fractions), combined with concurrent and adjuvant chemotherapy using temozolomide (TMZ) (142). Invariably, however, the disease recurs. No standard of care is established in recurrent or progressive GBM but treatments typically include supportive care, re-irradiation, surgery, systemic therapies, and combined modality therapy (143). Depending on the treatment and several risk factors, the median survival is around 9-14 months and only 5% of the patients are still alive at 5 years from the diagnosis (144). Notwithstanding ongoing efforts to test new curative approaches such as immunotherapy (145) and CAR-T cell therapy (146,147), nearly 200,000 people continue to die from GBM every year (148).

RT is often used as salvage treatment in recurrent GBM but the dose is limited by the risk of severe side effects (149). Proton therapy can spare more healthy tissue thanks to the Bragg peak, and indeed clinical trials have reported a reductions in grade >1 toxicity (150) and reduced lymphopenia (151) in GBM patients treated with protons vs. X-rays. However, the prognosis remains bleak. The only heavier particle currently used in the clinics is carbon, but clinical data from Japan (25) and Germany (152) had disappointing results for GBM patients. As noted above, the LET of C-ions is simply not high enough to overcome hypoxia (Fig. 2). Oxygen partial pressure in GBM is 2-3 times lower than in the normal brain (153,154). Hypoxia is directly responsible for GBM invasion and aggression, leading to poor patient outcomes (155–158).

The HI-FLASH project (https://www.gsi.de/HI-FLASH) will test $^{20}Ne^{10+}$-ions in FLASH conditions in a rat model of GBM, specifically the syngeneic F98/Fischer rat that is generally considered the best animal model of human GBM (159–161). The F98 tumour is weakly immunogenic, highly infiltrative, and very lethal (median survival 28 days, 100% death within 40 days (162)), just like a human GBM. The projectile for HI-FLASH will be $^{20}Ne^{10+}$-ions, as discussed above the most promising very heavy ions for hypoxic tumors. A recent experiment in Japan in a mouse model of squamous cell carcinoma shows indeed that the fraction of hypoxic cells is greatly reduced after irradiation with Ne-ions compared to X-rays (163). HI-FLASH will compare neurotoxicity and survival in F98 GBM rats irradiated with $^{20}Ne$-ions (whole brain irradiation) at ultra-high (100 Gy/s; exposure time 200 ms) or conventional (2 Gy/min; exposure time: 5-10 min) dose rate. The details of the beamline are shown in Fig. 8. As shown in the figure, normal rats will be exposed in plateau to test cognitive deficits while animals carrying the tumour will be irradiated in the SOBP to measure F98 tumor control and survival.

Rat models of glioma or glioblastoma have been already irradiated with ultra-high dose rate electrons (164) or protons (165). FLASH had a neuro-protective effect compared to conventional dose-rate, but no difference in tumor control or survival was noted. Based on these data, HI-FLASH will also use proton beams to demonstrate the FLASH effect in our system and to test the superiority of high-LET Ne-ions vs. low-LET protons in improving animal survival. Moreover, considering that in clinical applications it will be necessary to fractionate the dose to reduce the toxicity, HI-FLASH will also test whether the FLASH effect and the tumor control are maintained in a hypofractionated regime (3 daily fractions), and whether Ne-ions improve re-oxygenation (163).

**HEAVY ION IMAGING BY IN-BEAM PET**

An additional advantage of neon ions stems from nuclear fragmentation of the primary beam that leads to the production of positron emitting isotopes ([19]Ne) that can be visualized by positron emission tomography (PET). Online beam visualization is essential for reducing range uncertainty (166) and achieve real-time adaptive particle therapy (167). The use of [19]Ne-ion beams as a probe to check the range of [20]Ne-ion therapeutic beam was already explored during the LBNL trial (168) using a modified PET detector. We have recently demonstrated the feasibility of treating tumors with radioactive ion beams ([11]C), exploiting the increased intensity of the GSI accelerator and a dedicated PET detectors for small animals (169). A radioactive ion beam used for both treatment and imaging provides a much higher signal.to-noise ratio and a better alignment of the activity and dose peak (170). The half-life of [19]Ne is about 17.2 s, much shorter than [11]C and close to that of [10]C. We have shown that the short half-life of [10]C largely improves the spatial resolution of PET imaging compared to [11]C (171), and we therefore expect very high spatial sensitivity in imaging a [19]Ne-beam. Use of [19]Ne-ion beams for simultaneous treatment and imaging is another project under study.

**CONCLUSIONS**

The rationale for using heavy ions in radiotherapy lies in their biological advantages compared to protons. After the end of the LBNL trial over 30 years ago, only carbon ions have been used in clinical trials, with results that can be considered good but not matching the breakthrough expectations associated to early trials. To enhance the biological effectiveness, particularly for hypoxic tumors but also for eliciting other specific biological effects (such as a stronger immune response), the therapy should really exploit high-LET particles, and at therapeutic energies this is only possible with heavier ions. Clinical trials in Europe and Japan are considering using oxygen at least as a boost in multi-ion treatment. We propose neon ions as the best candidate to overcome hypoxic spatial heterogeneity in tumors. To reduce toxicity, minibeam and FLASH radiotherapy can be used. These two innovative approaches are being explored at high-energy accelerators both in Europe (GSI) and Asia (QST) in animal models. Range uncertainty can be reduced using in-beam PET visualization of the neon beam. We hope that these results will revive "Tobias' dream" of using very heavy ions in therapy.

**ACKNOWLEDGMENTS**

The HI-FLASH experiment is supported by the European Research Council (ERC) Adavanced Grant 2025 under Grant Agreement n. 101198395. Neon MBRT experiments  were supported in part by the Research Project Heavy Ions at NIRS-HIMAC, Particle Therapy Co-Operative Group (PTCOG) funding projects 2019 and from CNRS International Emerging actions (2020-2022). HADRONMBRT at GSI is supported by the Helmholtz Association. We thank Dr. Martina Moglioni and Dr. Daria Boscolo for their help with Fig. 4 and Fig. 8, respectively. Y.P. warmly thanks the support and collaboration at QST,  in particular Dr. Hirayama, Dr Shimokawa, Dr Inaniwa and Dr. Matsufuji.

Figure captions

**Figure 1.** Typical curves describing OER and RBE vs. LET in mammalian cells. A similar plot is found in ref. (9) based on the LBNL pre-clinical experiments in T1 cells. These curves were built from more recent data from Tinganelli et al. (22) using the curve at 0% oxygen concentration for Chinese hamster ovary cells, and a fit of the data in the comprehensive database for cell killing of mammalian cells with multiple ions at GSI PIDE (https://www.gsi.de/bio-pide) (172). Grey shaded area shows the LET range corresponding to the desirable values or OER and RBE in the target.

**Figure 2.** A comparison of dose-average LET in a single beam SOBP in a target tumour volume of 5x5x5 cm$^3$ (gray area) centered at 5 cm depth in water. Simulation by FLUKA code (base data) and TRiP98 (planning software) of H-, C-, and Ne-ions (orange, green and magenta lines, respectively). The yellow line shows 100 keV/$\mu$m, the minimum desirable LET in the SOBP. See Fig. 3 for a real treatment plan in GBM using two opposite beams.

**Figure 3.** Treatment planning of a GBM patient using opposite beams of carbon or neon ions. Treatment plan was calculated with TRiP98 using the LEM-IV RBE tables with $\alpha/\beta$ = 2 Gy. The dose distribution (**A**) shows that the two ions produce a similar conformal distribution on the target (clinical target volume, CTV, purple contour). CTV is much bigger than the visible cancer (gross tumour volume, GTV; black contour) owing to the highly diffusive nature of GBM that requires large margins. The LET distribution (**B**), is, however, much higher for Ne-ions, reaching the values above

>100 keV/μm in more than 90% of the CTV, as shown in the LET-volume histogram (**C**). CT image taken from the Cancer Imaging Archive (173).

**Figure 4.** Bragg curves in water calculated with the FLUKA Monte Carlo code in water for different ions at the same range (18 cm in water). **(A)** Dose normalized to the Bragg peak region. **(B)** Dose normalized to the entrance.

**Figure 5.** Example of multi-ion treatment at QST in Japan to increase the LET in the gross tumour volume in a pancreatic cancer case. **(A)** Dose distribution of carbon-ion radiotherapy. **(B)** LET distribution of the same case. Note overlying LET distribution near organs-at-risk. **(C)** The LET of the same case combining helium, carbon, and oxygen-ion treatments using the intensity modulated composite particle therapy system. Figure from ref. (40) reproduced under the terms of the Creative Commons Attribution License (CC BY).

**Figure 6.** Histology evaluation scoring results showing mean values and standard deviations comparing Neon MBRT versus conventional Neon broad-beam irradiation of C57BL/6J mice hind

limb. Figure from ref. (75) reproduced under   the terms of the Creative Commons Attribution License (CC BY).

**Figure 7.** Beamline in the GSI Cave A for FLASH experiments and close-up view of the 3D printed range modulator for unform 3D irradiation in the SOBP. Arrows point to beam monitor (blue), range shifter (red), range modulator (white) and target (green).

**Figure 8.** The HI-FLASH entire beamline corresponding to the photograph in Fig. 7.

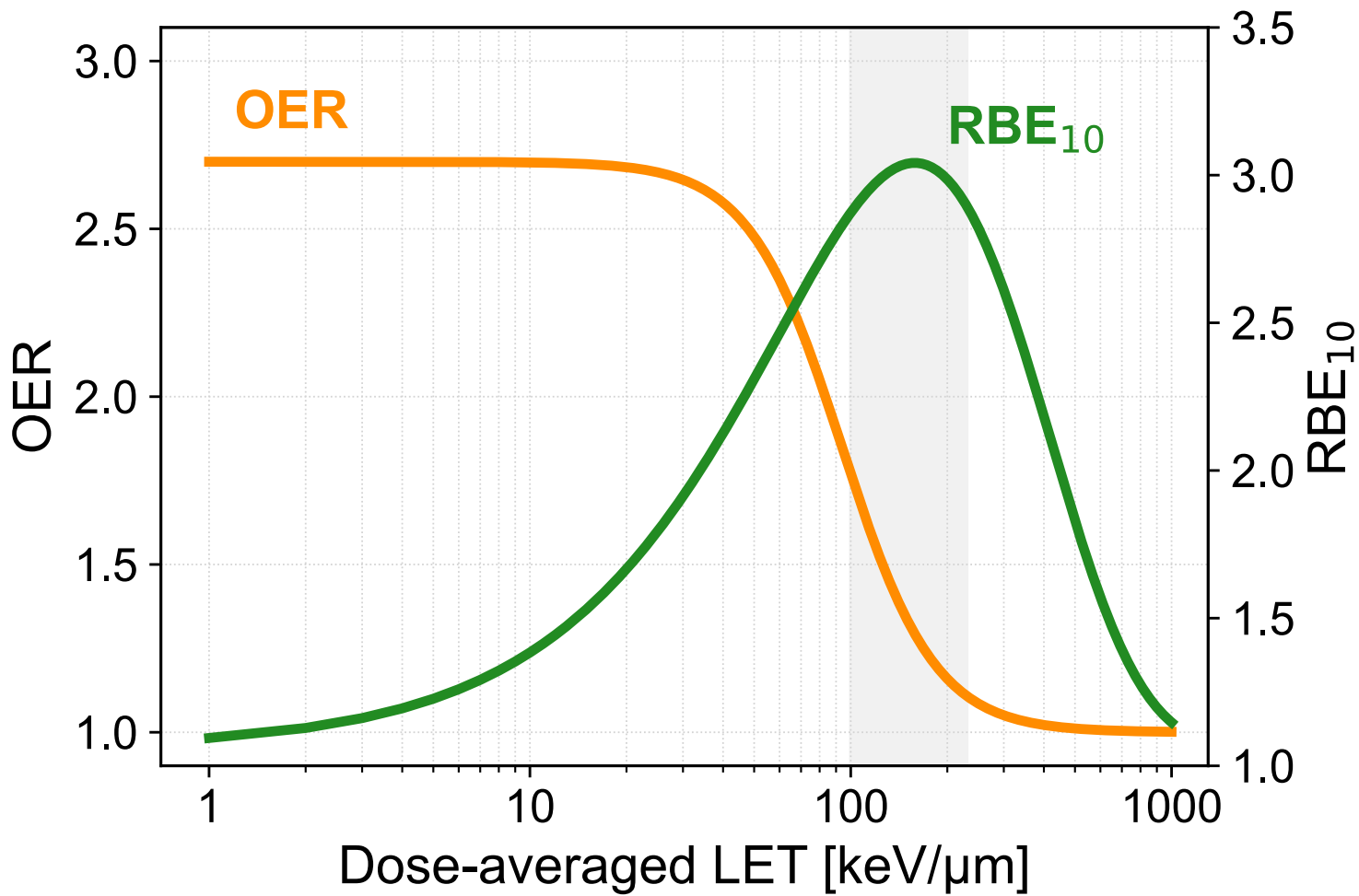

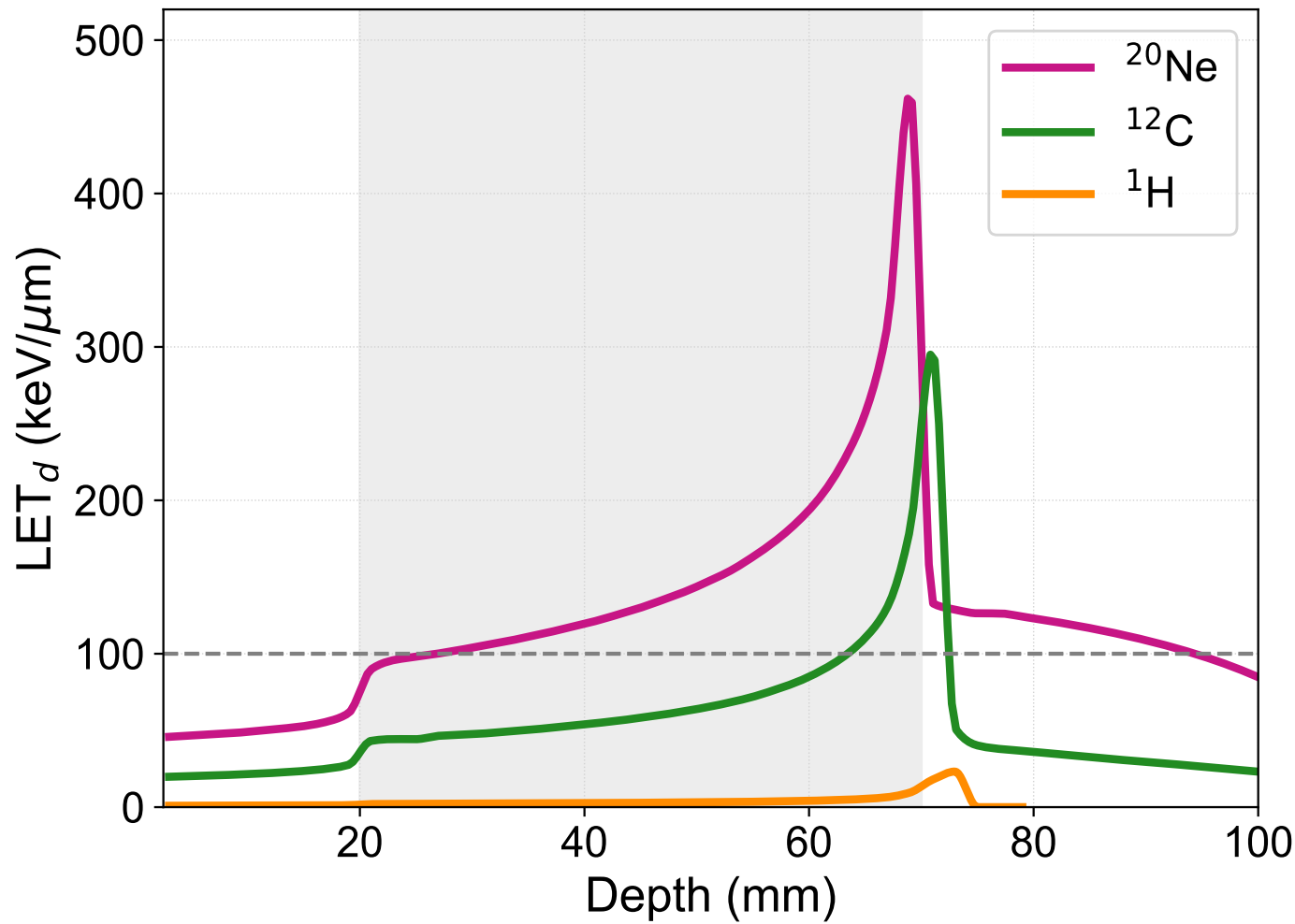

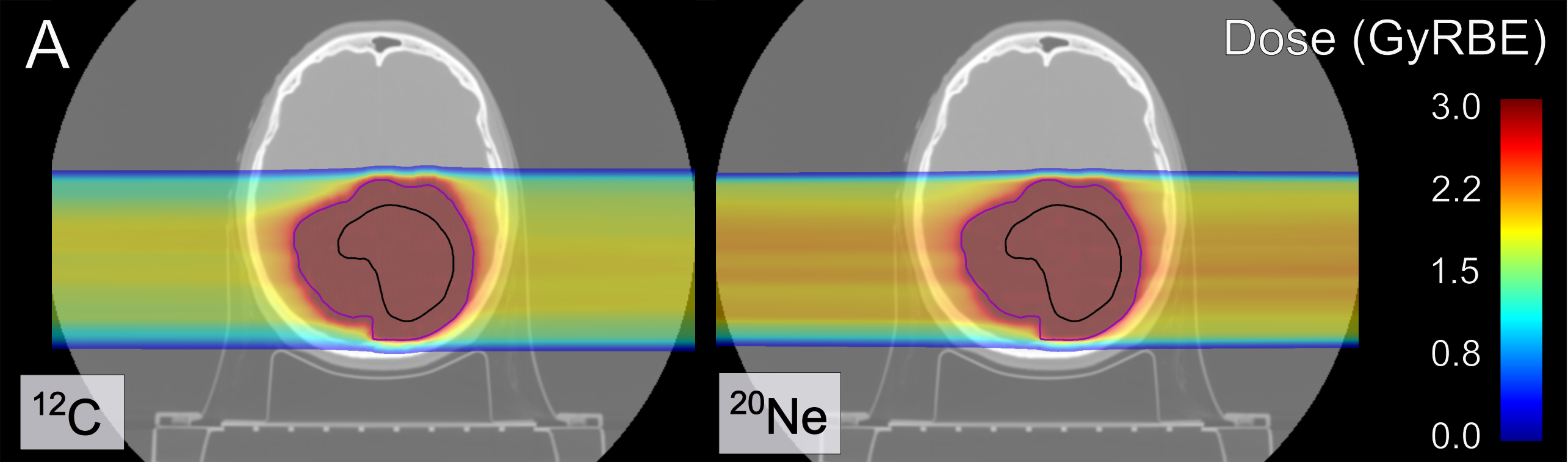

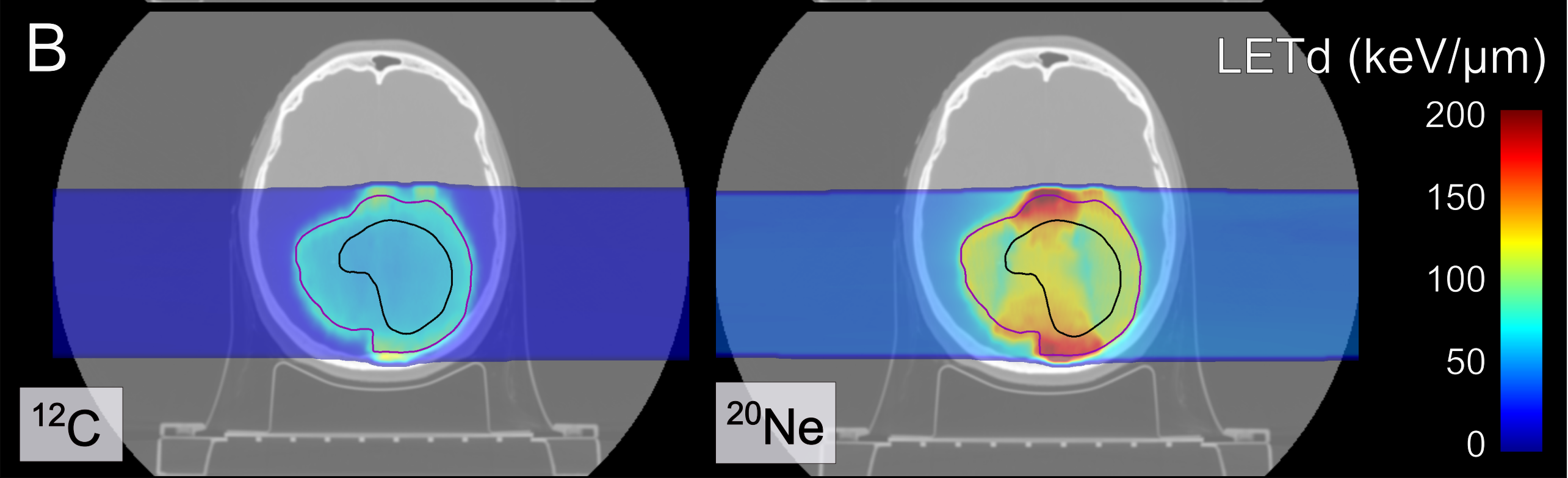

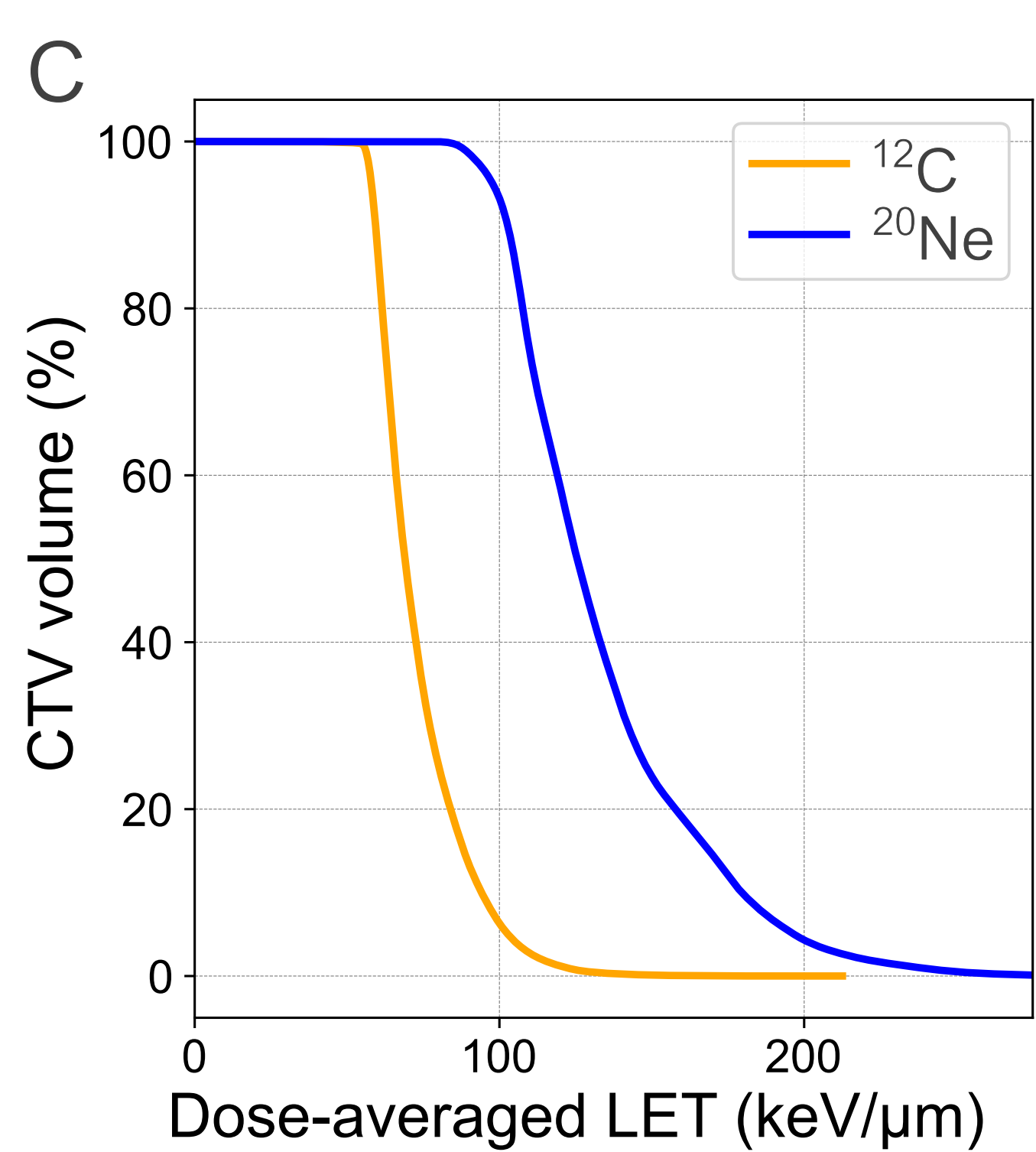

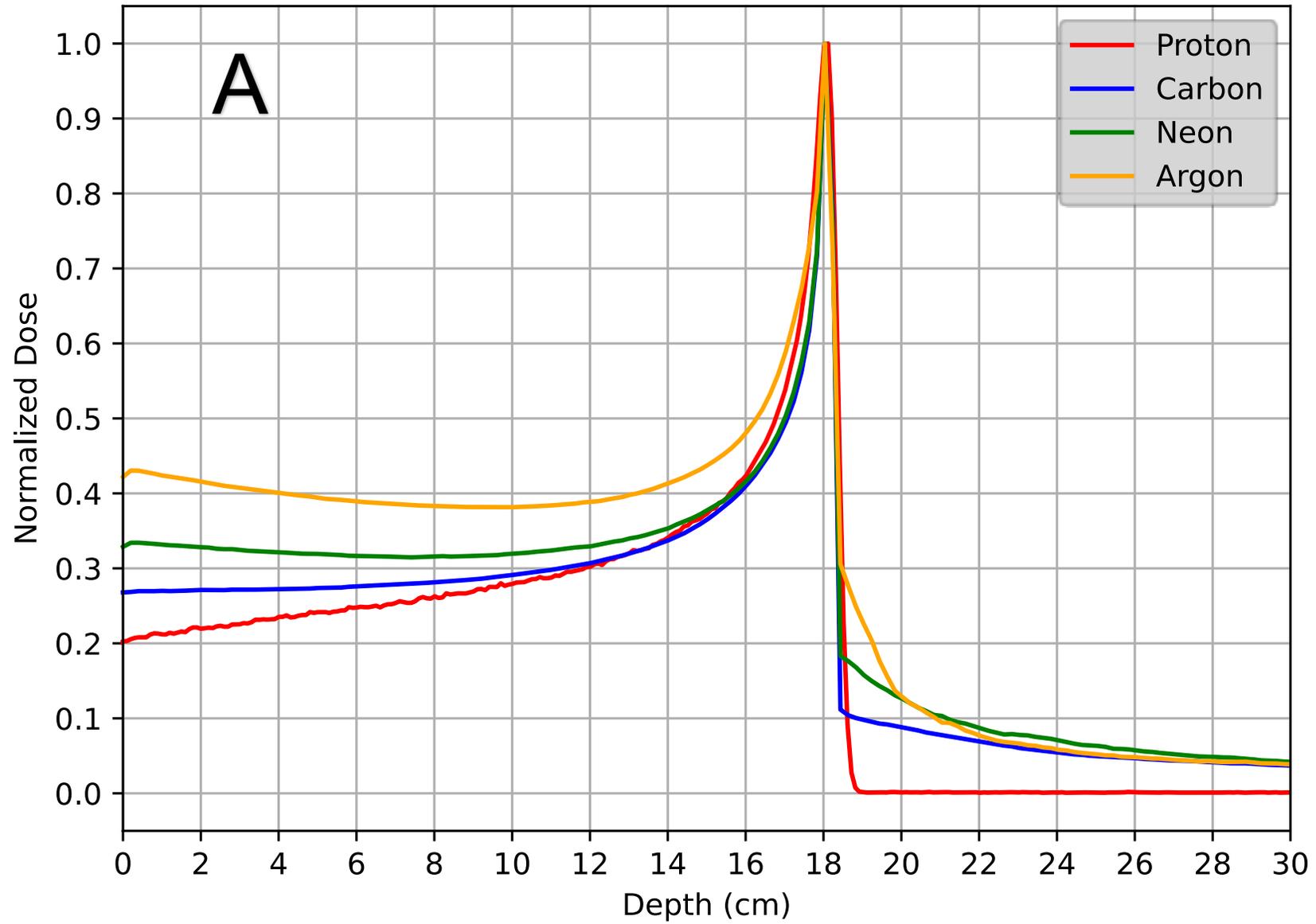

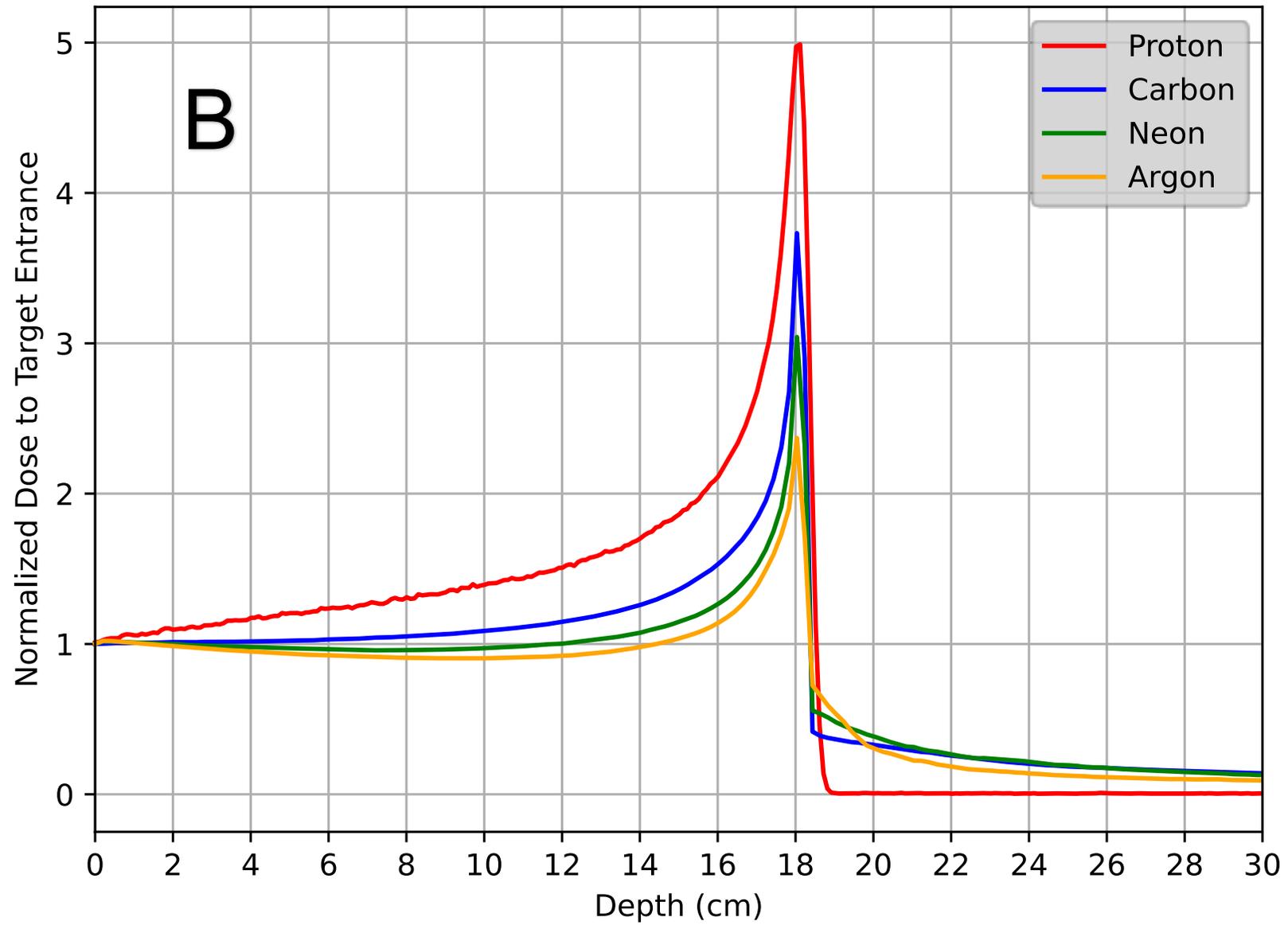

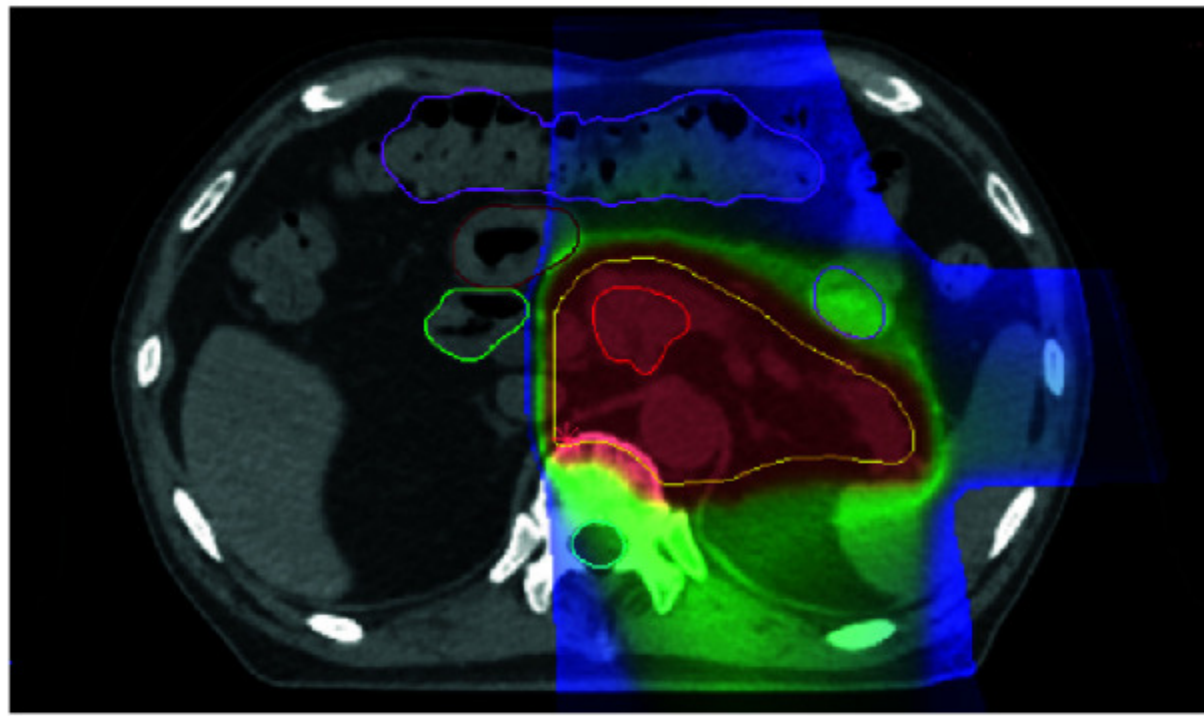

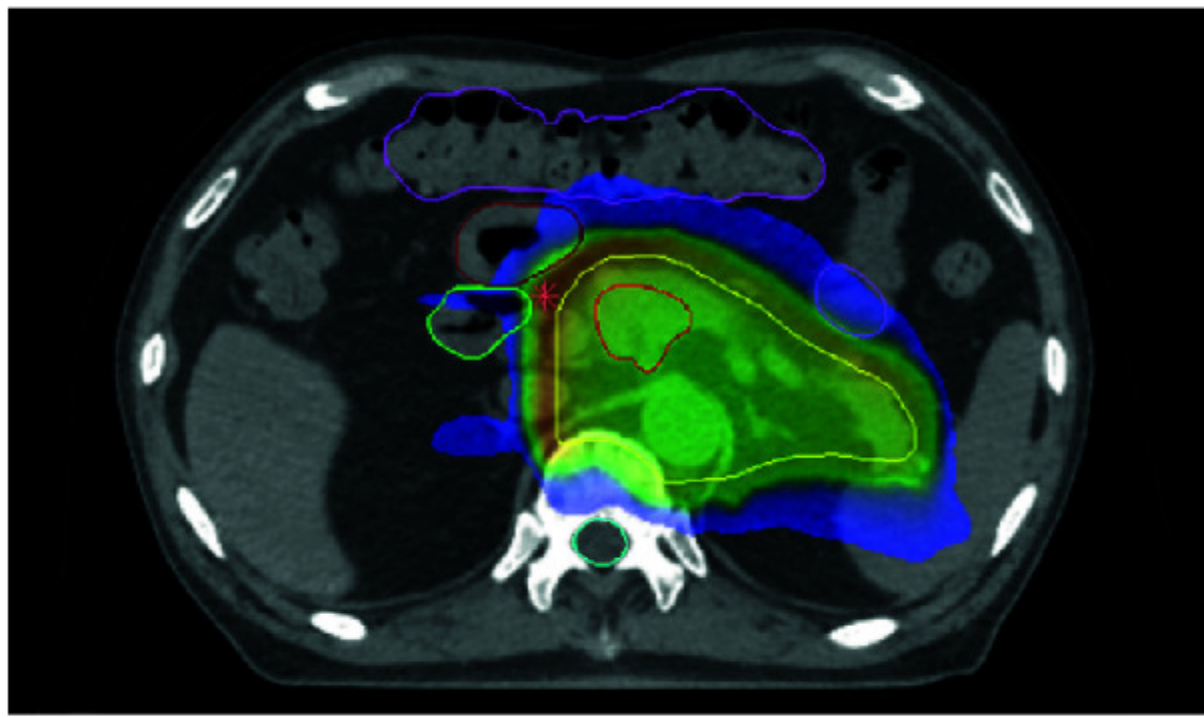

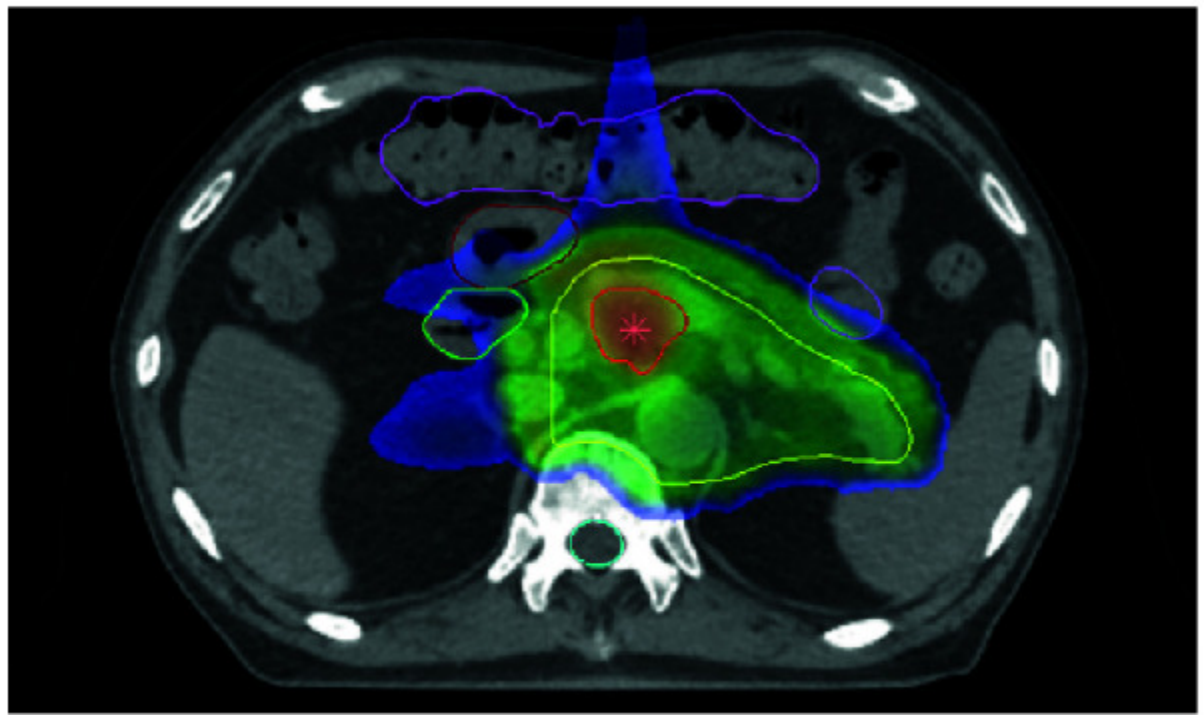

# Histopathology Evaluation Scoring Resuts

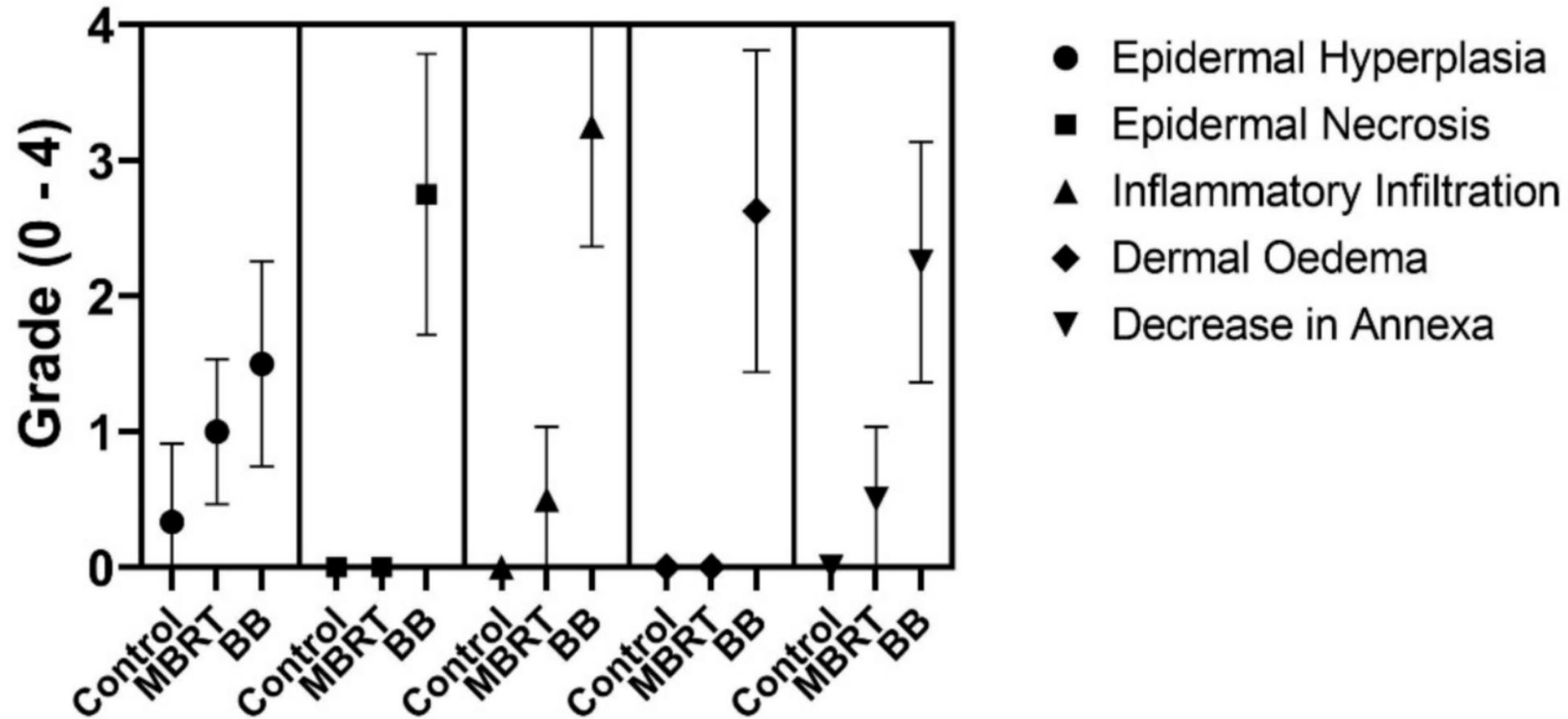

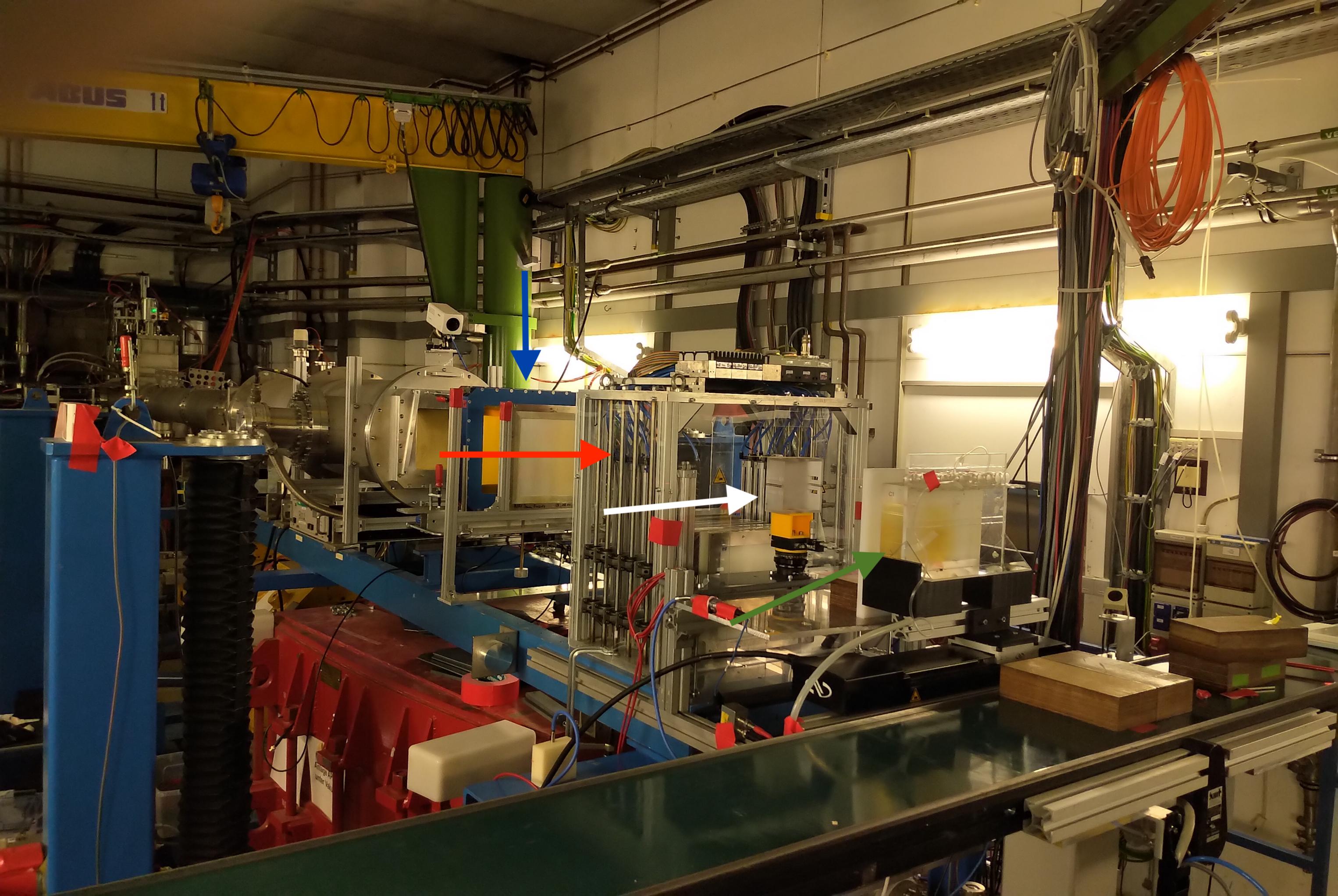

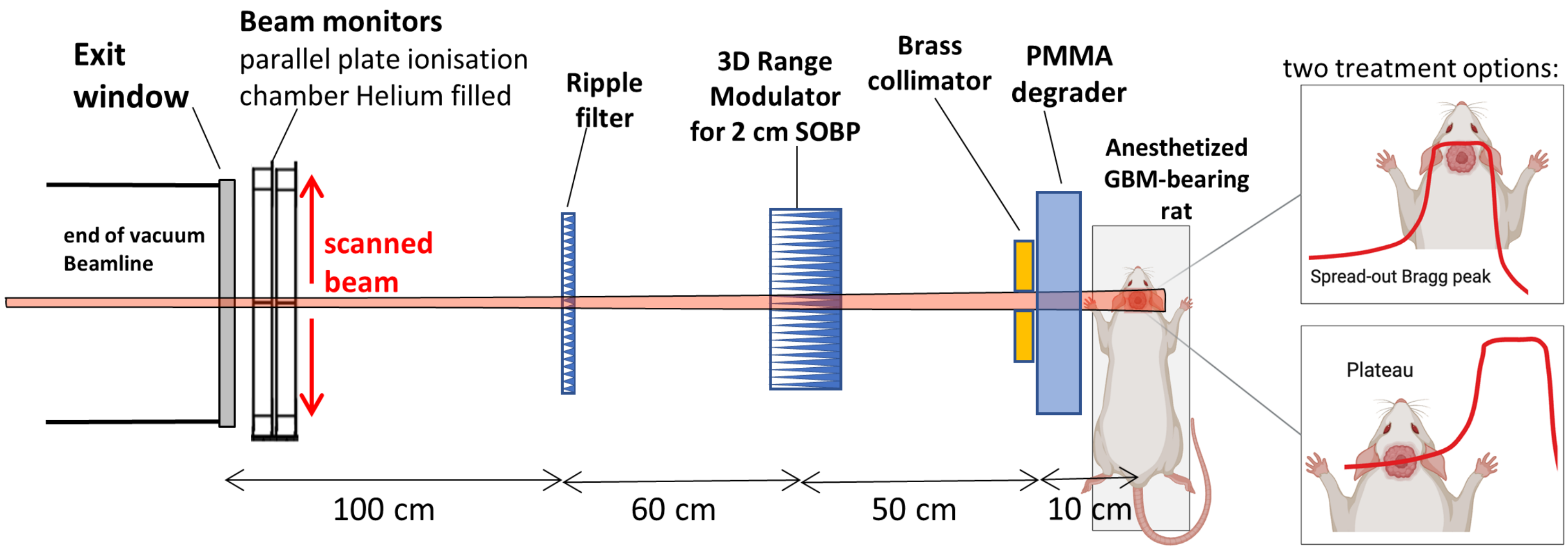